\documentclass[twocolumn,showkeys,aps,nofootinbib,superscriptaddress]{revtex4}

\usepackage[utf8]{inputenc}
\usepackage{graphicx} 
\usepackage{color}
\usepackage{amssymb}
\usepackage{amsmath}
\usepackage{amsfonts}
\usepackage{hyperref}
\usepackage{xcolor}
\usepackage{xfrac}

\usepackage{upgreek}
\usepackage{physics}

\begin{document}

\title{Search for the $1/2^+$ intruder state in $^{35}$P}
\date{\today}
\author{M.~Salathe}
\email{msalathe@lbl.gov}
\affiliation{Nuclear Science Division, Lawrence Berkeley National Laboratory, Berkeley, CA 94720, USA}


\author{H.~L.~Crawford}
\affiliation{Nuclear Science Division, Lawrence Berkeley National Laboratory, Berkeley, CA 94720, USA}

\author{A.~O.~Macchiavelli} 
\affiliation{Nuclear Science Division, Lawrence Berkeley National Laboratory, Berkeley, CA 94720, USA}

\author{B.~P.~Kay}
\affiliation{Physics Division, Argonne National Laboratory, Argonne, IL 60438, USA}

\author{C.~R.~Hoffman}
\affiliation{Physics Division, Argonne National Laboratory, Argonne, IL 60438, USA}


\author{A.~D.~Ayangeakaa}
\affiliation{Department of Physics and Astronomy, University of North Carolina at Chapel Hill, Chapel Hill, NC, 27599, USA.}
\affiliation{Triangle Universities Nuclear Laboratory, Duke University, Durham, NC 27708, USA.}

\author{C.~M.~Campbell}
\affiliation{Nuclear Science Division, Lawrence Berkeley National Laboratory, Berkeley, CA 94720, USA}

\author{R.~M.~Clark}
\affiliation{Nuclear Science Division, Lawrence Berkeley National Laboratory, Berkeley, CA 94720, USA}

\author{M.~Cromaz}
\affiliation{Nuclear Science Division, Lawrence Berkeley National Laboratory, Berkeley, CA 94720, USA}

\author{P.~Fallon}
\affiliation{Nuclear Science Division, Lawrence Berkeley National Laboratory, Berkeley, CA 94720, USA}

\author{M.~D.~Jones}
\affiliation{Nuclear Science Division, Lawrence Berkeley National Laboratory, Berkeley, CA 94720, USA}

\author{S.~A.~Kuvin}
\affiliation{Department of Physics, University of Connecticut, Storrs, CT 06269, USA}

\author{J.~Sethi}
\affiliation{Department of Chemistry and Biochemistry, University of Maryland, College Park, MD 20742, USA}

\author{M.~Wiedeking}
\affiliation{Department of Subatomic Physics, iThemba LABS, Somerset West 7129, South Africa}
\affiliation{School of Physics, University of the Witwatersrand, Johannesburg 2050, South Africa}

\author{J.~R.~Winkelbauer}
\affiliation{Los Alamos National Laboratory, Los Alamos, NM 87545, USA}

\author{A.~H.~Wuosmaa}
\affiliation{Department of Physics, University of Connecticut, Storrs, CT 06269, USA}


\begin{abstract}
  The excitation energy of deformed intruder states (specifically the 2p2h bandhead) as a function of proton number $Z$ along $N=20$ is of interest both in terms of better understanding the evolution of nuclear structure between spherical $^{40}$Ca and the Island of Inversion nuclei, and for benchmarking theoretical descriptions in this region. At the center of the $N=20$ Island of Inversion, the npnh (where n=2,4,6) neutron excitations across a diminished $N=20$ gap result in deformed and collective ground states, as observed in $^{32}$Mg. In heavier isotones, npnh excitations do not dominate in the ground states, but are present in the relatively low-lying level schemes.  With the aim of identifying the expected 2p2h$\otimes\mathrm{s}_{1/2^+}$ state in $^{35}$P, the only $N=20$ isotone for which the neutron 2p2h excitation bandhead has not yet been identified, the $^{36}$S(d,$^3$He)$^{35}$P reaction has been revisited in inverse kinematics with the HELical Orbit Spectrometer (HELIOS) at the Argonne Tandem Linac Accelerator System (ATLAS). While a candidate state has not been located, an upper limit for the transfer reaction cross-section to populate such a configuration within a 2.5 to 3.6\,MeV energy range, provides a stringent constraint on the wavefunction compositions in both $^{36}$S and $^{35}$P.

\end{abstract}


\maketitle
\section{Introduction}
\label{sec:intr}

The nature of shell structure of nuclei and its evolution with increasing neutron-proton asymmetry remains a fundamental question in nuclear structure research~\cite{Sorlin2008}. At the valley of $\beta$-stability, the $N = Z = 20$ shell closures are robust, and $^{40}$Ca is considered a doubly magic spherical nucleus, although deformed core-excited states have also been known for some time~\cite{Ber1966, Fed1969}.
However, it is also now well-known that as protons are removed from the $sd$-orbitals below Ca, the monopole shifts induced in the neutron single-particle levels effectively reduce the separation between the $\nu d_{3/2}$ and the $\nu f_{7/2}$ orbitals. This erosion of the $N=20$ $sd-pf$ shell gap, together with pairing and quadruple correlations, lowers the energetic cost for neutron pair excitations across the shell gap to the extent that multi-particle multi-hole configurations (e.g. 2p2h, 4p4h) become energetically favored. In the Island of Inversion centered around the neutron-rich Ne, Na, and Mg isotopes with $N\approx20$, collective and deformed ground states have been observed, and are attributed to a dominant contribution of these deformation-driving neutron-pair excitations to the ground-state wavefunction. 

Neutron particle-hole $sd-pf$ cross-shell intruder configurations do not dominate the ground state configurations in the heavier $N=20$ isotones ($Z>12$) but are still predicted to be present in the low-lying level scheme. The excitation energy of these intruder-dominated states, specifically the 2p2h bandhead along the the $N=20$ chain, provides information on the evolution of the $sd-pf$ shell gap.  Reproduction of the experimental trend in bandhead energy thus is a stringent test of theoretical descriptions in this region, particularly in terms of both cross-shell excitations and quadrupole correlations. However, measurements are sparse. 

\begin{figure}[tbhp!]
\centering 
\includegraphics{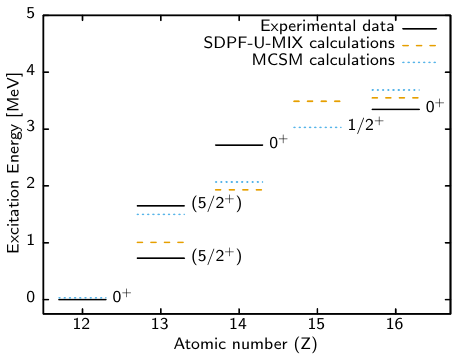}
\caption{Experimental (solid black lines) and calculated (dashed orange lines and dotted blue lines) 2p2h bandheads for the $N=20$ isotones between $Z=12$ and $Z=16$.  The orange dashed lines represent shell-model calculations performed with the SDPF-U-MIX effective interaction~\cite{Rotaru2012, Caurier2014, Valiente2018}, while the blue dotted lines are the results of MCSM calculations \cite{Utsuno2001, Wiedeking2008, Tripathi2008}.  The black solid lines represent data from Refs.~\cite{Mittig2002,Mutschler2016,Rotaru2012,Olness1971}.}
\label{fig:bh}
\end{figure}

The current state of affairs is summarized in Fig.~\ref{fig:bh}, with the evolution of the (tentative) experimentally determined 2p2h excitations along the $N=20$ isotones above Mg shown alongside theoretical predictions based on two different shell-model approaches. The calculated excitation energies for the lowest 2p2h-dominated state based on large-scale shell-model calculations with the SDPF-U-MIX effective interaction~\cite{Rotaru2012, Caurier2014, Valiente2018} are shown in the orange dashed lines, while the predictions of Monte-Carlo Shell Model (MCSM) calculations are shown in the blue-dotted lines~\cite{Utsuno2001, Wiedeking2008, Tripathi2008}. The solid black lines in Figure~\ref{fig:bh} represent the current best experimental candidate for the 2p2h bandhead in each $N=20$ isotone~\cite{Mutschler2016,Mittig2002,Rotaru2012,Olness1971}. Based on this figure, it is clear that while the general trend in behavior of the intruder states is well described by the available state-of-the-art shell model calculations, there remain discrepancies and important opportunities for refinement. Indeed, comparison of both level excitation energies and inferred wavefunction composition can be used to inform and improve model descriptions.

Following the initial $^{30}$Mg(t,p) measurement of Wimmer \textit{et al.}~\cite{Wimmer2010}, the $^{32}$Mg ground state was described as having a predominant intruder configuration.  This came into question briefly in the context of a two-level mixing model~\cite{Fortune2011}, but the $^{32}$Mg ground-state is now robustly described as having only very weak ($\approx4\%$) contributions from the 0p0h configuration and roughly equal 2p2h and 4p4h contributions~\cite{Macchiavelli2016}. In contrast, the $^{34}$Si ground-state has been estimated to consist of $\approx89\%$ 0p0h configurations, thus leaving as little as $11\%$ to contributions from states with neutron excitations~\cite{Rotaru2012}. The situation is experimentally less certain in $^{36}$S. The observation of the $0^+_2$ state in (t,p) reactions~\cite{Olness1971} and the absence of that state in (d,$^3$He) reactions~\cite{Gray1970} is a good indication that mixing is small and that the $0^+_2$ excited state is strongly dominated by neutron-pair excitations, while the ground state can be considered predominantly spherical. For the odd-$A$ nuclei there is only limited data available. In $^{33}$Al possible candidates have been proposed~\cite{Mittig2002,Mutschler2016}, however, the spin assignment of both the ground state and the candidate are yet to be confirmed. In $^{35}$P a candidate for the 2p2h bandhead still remains to be identified. A high-quality measurement clearly identifying the 2p2h bandhead in an odd-A $N=20$ isotone would provide an important confirmation for modern shell-model descriptions in this region of the nuclear chart. Moreover, a measurement of spectroscopic factors of the deformed states will allow a critical comparison to the theoretical wave functions.

In the case of $^{35}$P, the removal of a proton in the $^{36}$S(d,$^3$He) reaction will only populate the 2p2h state if there is non-zero mixing between the $^{35}$P ground state and the first 2p2h excitation and therefore significant overlap in the wave functions of these states. Previous investigations of this reaction, performed in the 1980s, did not observe any candidates for the 2p2h bandhead \cite{Thorn1984,Khan1985}. However, large background due to $^{12}$C contaminants in the $^{36}$S target dominated the $^3$He particle spectra of these experiments in the energy region between 3.0 and 3.5~MeV where the bandhead would be expected (MCSM calculations predict the bandhead at 3.03~MeV, as shown in Figure~\ref{fig:bh}). Thus, these experiments could not be conclusive on the observation or lack thereof for the intruder state.  

We report here on a recent measurement of the $^{36}$S(d, $^{3}$He)$^{35}$P reaction studied in inverse kinematics with the HELical Orbit Spectrometer (HELIOS) \cite{Lighthall2010}.  This approach offers a clean measurement free of the background observed in normal kinematics experiments. Thus, while we did not observe any state consistent with the 2p2h bandhead, we are able to set an upper limit on the spectroscopic factor as a function of the energy of the expected intruder state.  This in turn provides a constraint on the 0p0h and 2p2h content of the wavefunctions.  


\section{Experiment}
\label{sec:exp}
\begin{figure*}
\centering \includegraphics{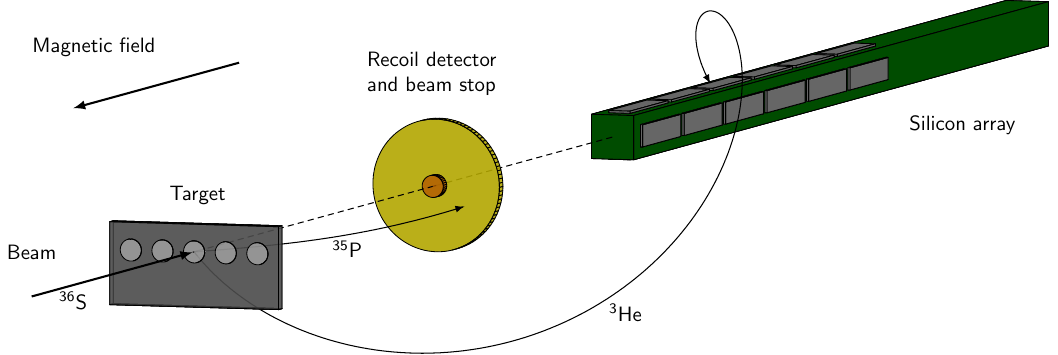}
\caption{\label{fig:setup}Schematic representation of the experimental setup installed in HELIOS. The incoming beam ($^{36}$S) hits a deuterated-plastic targets placed in the center of the solenoid. The heavy reaction products are measured with the recoil detector or stopped in the (inactive) beam stop. The light particles ($^3$He) spin in the magnetic field until they hit the silicon array installed behind the recoil detector.}
\end{figure*}

\begin{figure}
\centering \includegraphics{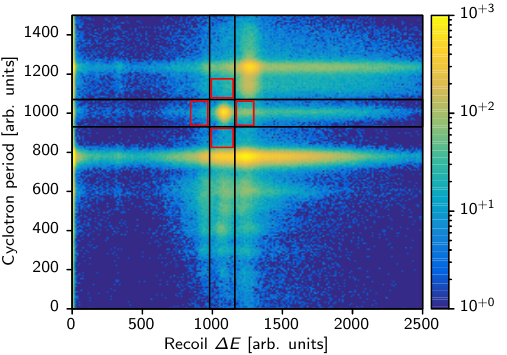}
\caption{\label{fig:gates}A representation of the two main analysis cuts used to filter events. The cyclotron period is proportional to the mass over charge ratio of the light particle and the area between the two horizontal lines is the location of $^3$He particles. The recoil energy loss is proportional to the heavy particles $Z$ and the two vertical lines select $Z=15$. The color-scale represents the number of particles in the region below 6\,MeV excitation energy. The red squares indicate the areas used to estimate the background component in the center gate. The background counts were weighted by a factor of $\sfrac{1}{3}$ to compensate for the larger coverage of the background gate.}
\end{figure}

The structure of $^{35}$P has been studied in inverse kinematics with HELIOS \cite{Lighthall2010} located at the Argonne National Laboratory. The Argonne Tandem Linac Accelerator System (ATLAS) provided a stable $^{36}$S beam at 15.3\,MeV$/A$. The beam impinged on a range of deuterated-plastic targets (81, 127, 529\,$\mu$g$/$cm$^2$ thicknesses) located in the bore of the HELIOS solenoid magnet (operated at a magnetic field strength of 2.85\,T). Both the $^3$He ions and the $^{35}$P were emitted at forward, near on-axis lab angles. As illustrated in Fig. \ref{fig:setup}, the $^3$He ions spiral in the magnetic field and are collected on a position sensitive silicon array, placed along the beam axis. Depending on the emission angle and energy, the $^3$He particles intercept the silicon array at different positions. The silicon detectors were located between 58-93\,cm from the target, corresponding to a maximal angular range of 10-50$^\circ$ in the center-of-mass frame.  Due to the poor resolution obtained in some of the silicon detectors, only a subset were included in the present analysis. 

The energy loss of $^{35}$P and scattered $^{36}$S particles, as well as background recoils from fusion-evaporation reactions, was measured with a 65\,$\mu$m thick silicon detector (recoil detector) installed between the target and the silicon array. The information was used to select $Z=15$ recoils; the observed pulse-height distribution and the $Z=15$-gate are represented in Fig.~\ref{fig:gates} on the x-axis. A beam blocker with a $\approx$10\,mm diameter was placed on the recoil detector, centered on the beam axis, to limit the overall rate.

The cyclotron period of the outgoing ions can be identified with respect to the radio frequency (RF) structure of the accelerator, for which the beam is delivered in bunches 1-2\,ns wide every $82.47$\,ns. The time delay between the ATLAS RF and detection of an ion in the silicon array is proportional to the mass of the particle hitting the array, divided by its charge. This measure of the cyclotron period allows for selection of $^3$He particles detected on the silicon array. The observed cyclotron period and the $^3$He-gate is shown in Fig.~\ref{fig:gates} on the y-axis. The $^3$He-gate along with selection of $Z=15$ heavy recoils allowed the necessary rejection of background in the excitation energy region where the $^{35}$P states are detected and are the main cuts applied to the data. Fig.~\ref{fig:gates} also shows the four nearest neighbor gates symmetrically distributed around the main gate that were used to estimate backgrounds.

The energy of $^3$He ions measured on a given silicon detector is related to the location at which the particle hits the detectors. This relationship between energy and return distance $z$ was described in Ref.~\cite{Wuosmaa2007} and is:
\begin{equation}
E_\mathrm{lab}=E_\mathrm{cm}-\frac{1}{2}mV_\mathrm{cm}^2+(\frac{mV_\mathrm{cm}}{T_\mathrm{cyc}})z.
\label{eq:ecm}
\end{equation}
The cyclotron time $T_\mathrm{cyc}$, the particle mass $m$ and the velocity of the center-of-mass frame with respect to the laboratory frame $V_\mathrm{cm}$ are all constants for a given experiment. Thus, for a constant $E_\mathrm{cm}$ there is a linear relationship between the observed energy and interaction location. Ballistic effects within some of the detectors add distortions that depend on the location at which the particles hit a given detector. The $^{36}$S(d,$^3$He)$^{35}$P reaction populates mainly the ground state and the excited $5/2^+_1$ state at 3860\,keV. The energy dependence on the position was removed based on a polynomial fit to the ground state. The individual detectors were then gain matched according to the known energies of these two states. Furthermore, a residual shift of the three peaks observed above 3860\,keV was removed by matching these peaks to the literature values through a linear fit.

\section{Results}

\begin{figure}
\centering \includegraphics{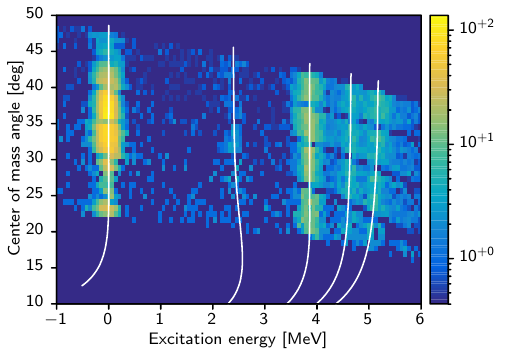}
\caption{\label{fig:zvsen}Center-of-mass angle (in degrees) vs. excitation energy for all events corresponding to detection of $^{3}$He and a $Z=15$ heavy fragment. The color-scale shows the number of observed events, the white lines represent the location of the states calculated from simulations of the setup.}
\end{figure}
\begin{figure}
\centering \includegraphics{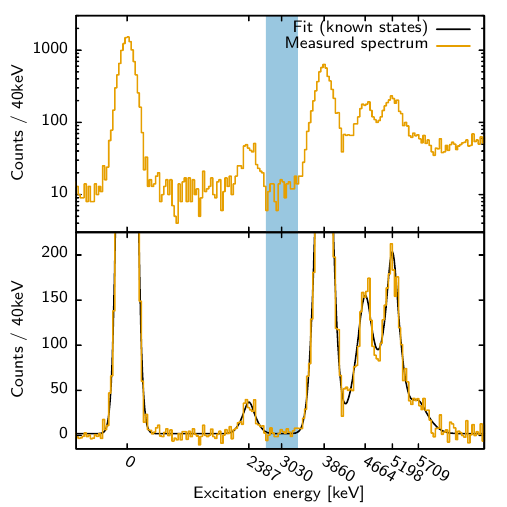}
\caption{\label{fig:enehis}The excitation energy measured in the $^{36}$S(d,$^3$He)$^{35}$P reaction. The blue area indicates the region of interest for a potential 2p2h bandhead (bounded by measurements in neighboring isotones).  The vertical axis of the top panel is logarithmic.  The bottom panel shows a linear scale, with the background defined in Fig.~\ref{fig:gates} subtracted.}
\end{figure}
\begin{table*}
  \begin{tabular}{c|c|c|c|c|c|c|c}
  \hline \hline
    State & Energy & Peak centroid & Peak Counts & $C^2S$ & $C^2S $ & $C^2S$ & $C^2S$ \\
    $I^\pi$ \cite{Chen2011, Mutschler2016b}  & [keV] \cite{Chen2011, Mutschler2016b} & [keV] &  &\cite{Thorn1984}  & \cite{Khan1985} & \cite{Mutschler2016b} &This work \\
    \hline
    $1/2^+$ & 0              & $0\pm1$     & $10478\pm105$ & 2.0 & 2.0 & 2.0 & 2.0 \\ 
    $3/2^+$ & $2386.6\pm0.5$ & $2388\pm13$ & $278\pm25$  & --  &  0.4(1)        & 0.6(3) & $0.33 \pm 0.03 \mathrm{(stat.)} \pm 0.08 \mathrm{(syst.)}$ \\
    $5/2^+$ & $3859.9\pm0.5$ & $3860\pm2$  & $4817\pm75$ & 1.0(7) &   3.6(1)   & 2.5(1) & $2.92 \pm 0.06 \mathrm{(stat.)} \pm 1.04\mathrm{(syst.)}$ \\
    $5/2^+$ & $4664\pm3$     & $4666\pm9$  & $1758\pm81$ & 0.3(3)  &   1.3(3)   & 0.9(3) & $0.71 \pm 0.02 \mathrm{(stat.)} \pm 0.34 \mathrm{(syst.)}$ \\
    $5/2^+$ & $5198\pm10$    & $5202\pm8$  & $1767\pm113$  & 0.3(2) & 1.7(5)    & 1.4(5)&  $1.10 \pm 0.03 \mathrm{(stat.)} \pm 0.57\mathrm{(syst.)}$ \\
    ($1/2^-$)   &  $5709\pm20$ & $5706\pm38$ & $395\pm79$ & -- & -- & 0.19(15) & $0.23 \pm  0.02 \mathrm{(stat.)} \pm 0.05 \mathrm{(syst.)}$ \\
    \hline \hline
  \end{tabular}
  \caption{\label{tab:fit} An overview over the measured quantities and comparison with previous measurements. The spectroscopic factors have been normalized to 2 for  the ground state values. The peak centroids and  peak counts were derived from the background subtracted energy spectrum.}
\end{table*}
\begin{figure}
\centering \includegraphics{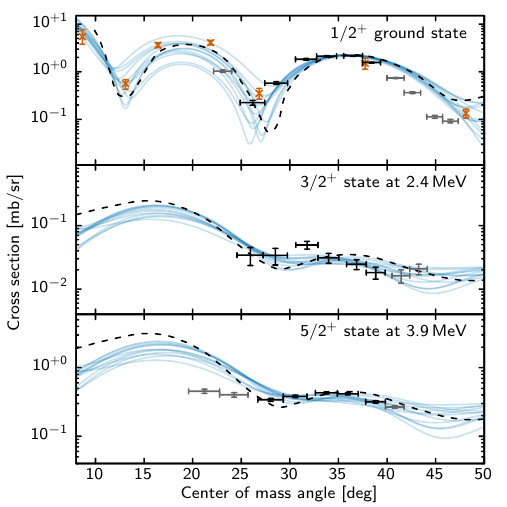}
\caption{\label{fig:angdis} Angular distributions for the three lowest energy states (black/gray error bars) are compared to DWBA calculations. Data and models are normalized to the ground state distribution of Ref.~\cite{Thorn1984} which is shown with orange crosses. The black dashed line is the DWBA calculation presented in Ref.~\cite{Thorn1984}. The blue lines illustrate PTOLEMY DWBA results using the possible combinations of optical potentials~\cite{Daehnick1980, An2006, Han2006, Trost1987, Liang2009, Pang2009, Xu2011}. The points marked with gray error bars were not used in the fit so that all states were fit over a similar angular range.} 
\end{figure}

The resulting event distribution as a function of center-of-mass angle and energy is shown in Fig.~\ref{fig:zvsen} and the projection onto the energy axis is given in the upper panel of Fig.~\ref{fig:enehis}. The lower panel of Fig.~\ref{fig:enehis} shows a background subtracted version of the energy spectrum, that is used to extract the background subtracted peak counts and the peak positions as summarized in Table~\ref{tab:fit}. For this purpose, the 6 most prominent states below 6\,MeV excitation energy, have been fit with a functional form that assumes constant background and two Gaussian distributions with identical centroids for each individual peak. The peak-height and width ratios between the two Gaussian distributions were required to be identical for all peaks. A pair of Gaussian distributions was used to accommodate the facts that peaks are composed of counts from multiple detectors of different resolutions and that a single Gaussian distribution did not describe the observed peak shape robustly. The fit was performed with a Poisson maximum-likelihood approach. The quoted uncertainties for the peak counts are statistical only. The peak resolutions (defined as the mean of the two Gaussian's $\sigma$ weighted by their respective counts) varied between 118\,keV (ground state) and 183\,keV (highest excitation energy). A state at 4494\,keV was observed in \cite{Mutschler2016b} and may account for the small excess in counts visible in the measured spectrum between the first and second $5/2^+$ states.  However, due to overlapping peaks in this region, this peak was not included in the fit. In the region of interest for a potential 2p2h bandhead candidate, marked blue in Fig.~\ref{fig:enehis}, no peak is observed above what would be expected from a flat background. Without subtraction, the background in the region of interest was estimated at 316$\pm$16\,counts$/$MeV.

As discussed previously, the position along the beam axis and the energy of the detected particle can be used to determine the emission angle in the center-of-mass frame \cite{Wuosmaa2007}, yielding the angular distributions shown in Fig.~\ref{fig:angdis} for the ground-state (top panel) and first two excited-states in $^{35}$P (middle and bottom panels). The angular distribution can be calculated through the distorted-wave Born approximation (DWBA). The relative scaling of the data to DWBA calculation is directly proportional to the spectroscopic factor. To derive relative spectroscopic factors, background subtracted data were weighted by their uncertainties and fit to the DWBA calculations from an earlier measurement conducted at a similar center-of-mass energy \cite{Thorn1984}.  As measurement of the beam current was not made with sufficient accuracy to calculate absolute values, the relative spectroscopic factors, listed in Table~\ref{tab:fit}, were normalized such that the ground state value is 2. The derived spectroscopic factors are in agreement with earlier measurements and were used to align data and DWBA calculation in Fig.~\ref{fig:angdis}. The DWBA calculation from the earlier measurement \cite{Thorn1984} used for extracting the spectroscopic factors are shown with dashed black lines in Fig.~\ref{fig:angdis} and data from that publication are shown (in orange) for the ground state. Furthermore, a variety of DWBA calculations performed with PTOLEMY \cite{Ptolemy2017} have been added to Fig.~\ref{fig:angdis}; the incoming particle (deuteron) optical potentials were taken from Refs.~\cite{Daehnick1980, An2006, Han2006} and the outgoing particle ($^{3}$He) potentials from Refs.~\cite{Trost1987, Liang2009, Pang2009, Xu2011}. We note that the angular coverage of the current results cover the second maximum for all states considered and that the magnitude of the absolute cross sections differ between the global parameterizations and those of Ref.~\cite{Thorn1984}. The relative spectroscopic factors were also determined for the PTOLEMY based DWBA calculation; the standard deviation between the different choices for optical potentials was used to estimate the systematic uncertainty listed in Table~\ref{tab:fit}. The spectroscopic factor for the 5709\,keV transition is exclusively based on PTOLEMY calculations as that transition was not observed in \cite{Thorn1984} and thus, it might be affected by different systematic effects.

\begin{figure}
\includegraphics[width=\linewidth]{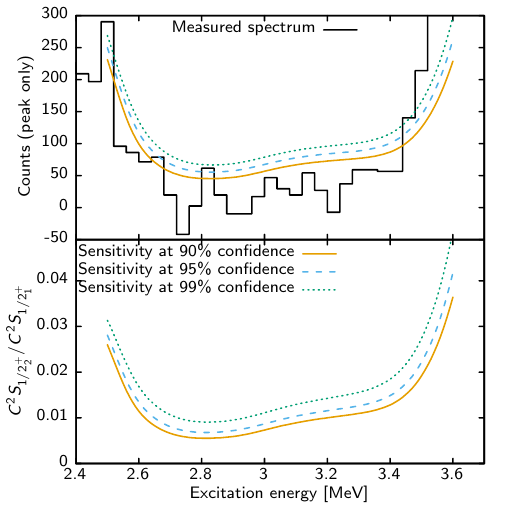}
\caption{\label{fig:sens} Top panel: number of counts in a peak in the region of interest that would be required to reject the null result (no additional peak present) with a given confidence. The background subtracted spectrum is shown to guide the eye. Bottom panel: Exclusion curves for the $C^2S$ ratio between the ground and first excited $1/2^+$ state.}
\end{figure}

Turning to the region of interest with respect to a potential 2p2h bandhead, the experimental sensitivity at a given energy was estimated as the maximum number of counts in a peak added to the statistical fluctuations, such that the minimized model distribution does not exceed a predefined confidence level (90\%) when being compared to the observed data. The additional peak was also made up of two Gaussian distributions and its resolution was fixed to a linear interpolated value between the two adjacent peaks resolutions. It was placed in the energy range between 2.5 and 3.6\,MeV and the remaining free parameters in the model found by minimizing the Poisson maximum likelihood of the model with respect to the spectrum without background subtraction. For simplicity, Pearsons $\chi^2$ was used to approximate the p-values of the Poisson maximum likelihood. The number of counts required for a possible observation are shown as a function of energy in the top panel of Fig.~\ref{fig:sens}. The bottom panel uses this information, together with DWBA calculation (based on Refs.~\cite{An2006,Pang2009}) to establish an upper limit for the $C^2S$ ratio between ground and excited state.

\section{Discussion}

\begin{figure*} 
\includegraphics[width=\linewidth]{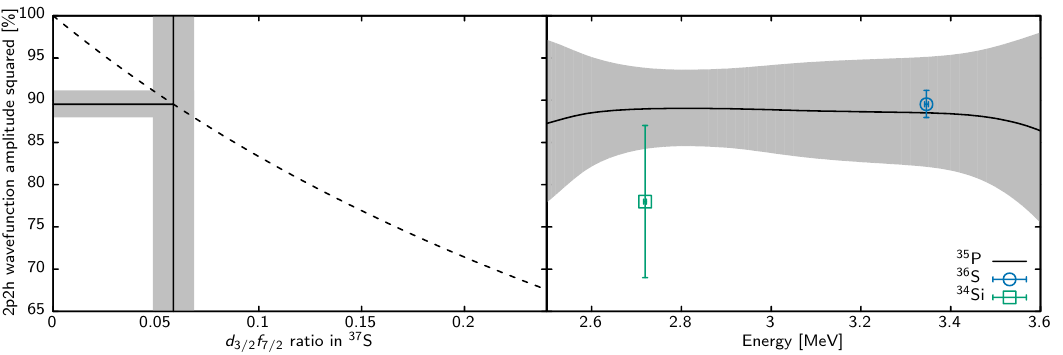} 
\caption{\label{fig:2by2} Sensitivity analysis in the 2x2 model. Left panel: Amplitude squared, $\alpha^2$, of the 2p2h component in the $0^+_2$ state of $^{36}$S, derived from the $^{36}$S(d,p)$^{37}$S reaction (black line and grey bands) and the $C^2S$ ratio from Eq.~\ref{eq:36Scsratio} (dashed line). Right panel: Lower limits on the 2p2h excitation amplitude, $A^2$, in $^{35}$P, derived from the experimental 90\% confidence sensitivity as a function of the expected energy of the excited state. The amplitude $\alpha^2$ is shown at the energy of the $0^+_2$ in $^{36}$S (blue circle) together with that for $^{34}$Si \cite{Rotaru2012} (green square).}
\end{figure*}

The impact of the upper limit for the ratio of the spectroscopic factors between a potential 2p2h bandhead in the region of interest and the ground state can be gauged by considering a simplified 2$\times$2 (two-state) mixing model.
Studies of the $^{36}$S(d,p)$^{37}$S reaction~\cite{ECKLE1989,PISKOR1984,Thorn1984} show the population of a $d_{3/2}$ hole in the ground state of $^{36}$S and can provide an assessment of the proportion of 2p2h excitations present in the $0^+_1$ state.
Consider that the ground state wave-function of $^{36}$S is described in a simple form as\footnote{The corresponding orthogonal $0^+_2$ state is
$\ket{0_2^+} = -\beta\ket{\mathrm{0p0h}}+\alpha\ket{\mathrm{2p2h}}$}: 
\begin{align}
\ket{0_1^+} =  (\alpha\ket{\mathrm{0p0h}}+\beta\ket{\mathrm{2p2h}})
\label{eq:36Swave}
\end{align}
 \noindent
 where $\ket{\mathrm{0p0h}} \approx d_{3/2}^4$ and $\ket{\mathrm{2p2h}} \approx d_{3/2}^2f_{7/2}^2$. The experimental ratio of the neutron spectroscopic factors for the population of the $7/2^-$ and $3/2^+$ in $^{37}$S in the $(d,p)$ reaction can be readily calculated from Eq.~\ref{eq:36Swave}:
 \begin{align}
\frac{C^2S_{3/2^+}}{C^2S_{7/2^-}}=\frac{1}{2}\left(\frac{\beta}{\alpha}\right)^2
\label{eq:36Scsratio}
\end{align}
\noindent
Fig.~\ref{fig:2by2} (left panel) shows the behavior of amplitude $\alpha^2$ as a function of this ratio. When compared with the average (and its standard deviation) obtained from the data in Refs. ~\cite{ECKLE1989,PISKOR1984,Thorn1984} we can determine $\alpha^2 = 89.5\pm1.6\%$ which, as anticipated, corresponds essentially to a 0p0h configuration for the ground-state of $^{36}$S.

Proceeding now to $^{35}$P the ground state $\ket{1/2_1^+}$ and the excited $\ket{1/2_2^+}$ can be described in the simple two-level model as:
\begin{align}
\ket{1/2_1^+} =  \left(A\ket{\mathrm{0p0h}}+B\ket{\mathrm{2p2h}}\right)\otimes\pi s_{1/2}
\label{eq:35Pgs}
\end{align}
\begin{align}
\ket{1/2_2^+} = \left(-B\ket{\mathrm{0p0h}}+A\ket{\mathrm{2p2h}}\right)\otimes \pi s_{1/2}
\label{eq:35Pes}
\end{align}
\noindent
and following from Eqs.~\ref{eq:36Swave}, \ref{eq:35Pgs} and \ref{eq:35Pes} we then estimate the ratio of spectroscopic factors as
\begin{align}
\frac{C^2S_{1/2_2^+}}{C^2S_{1/2_1^+}} \approx  \frac{(-\alpha B+\beta A)^2}{(\alpha A + \beta B)^2}
\label{eq:35Pcsratio}
\end{align}
\noindent 
It is interesting to note that because of the interference in the numerator of Eq. \ref{eq:35Pcsratio}, the stringent limits set by HELIOS (see Fig.~\ref{fig:sens}) with the non-observation of a candidate peak, can be applied to establish a meaningful limit on the values of the amplitude $A^2$ as shown in the right of Fig.~\ref{fig:2by2}, in the energy range expected for the location of the $1/2_2^+$ state.   Thus, the sensitivity analysis based on the 2$\times$2 model suggests the similarity between $^{35}$P and $^{36}$S in terms of the evolution of shape coexistence towards the center of the $N=20$ Island of Inversion centered at  $^{32}$Mg. 


\section{Conclusion}

In search of the 2p2h bandhead in $^{35}$P, the $^{36}$S(d,$^3$He)$^{35}$P reaction has been revisited in inverse kinematics with HELIOS. However, no candidate peak was observed in the expected region of interest between 2.5\,MeV and 3.6\,MeV. Based on studies of the $^{36}$S(d,p)$^{37}$S reaction~\cite{ECKLE1989,PISKOR1984,Thorn1984} and a 2$\times$2 model the 0p0h waveform amplitude of the $^{36}$S ground state was derived to be $89.5\pm1.6\%$. Based on this result, the non-observation of a candidate peak sets a tight lower limit on the 2p2h waveform amplitude for the (still-to-be-observed) $1/2^+_2$ intruder state in $^{35}$P.

Given the interference between the unperturbed $1/2^+$ states discussed above, it is not clear that an experiment with more statistic and higher sensitivity will result in a positive observation of the intruder state with the (d,$^3$He) reaction.  In this regard, a study of the $^{33}$P(t,p)$^{35}$P and  $^{37}$P(p,t)$^{35}$P reactions is suggested. In the former, stripping of 2 neutrons into the $fp$-shell naturally leads to 2p2h configurations in $^{35}$P; in the latter, these states can be populated by the pickup of 2 neutrons from the closed $sd$ shell. These experiments could be carried-out with the new spectrometer SOLARIS~\cite{Solaris} at FRIB, where re-accelerated beams of $^{33,37}$P of adequate intensity will be available on day one~\cite{Beams}.

\section*{Acknowledgment}

This material is based upon work supported by LBNL-LDRD funding under LDRD NS16-128, and by the U.S.\ Department of Energy, Office of Science, Office of Nuclear Physics under Contract No.~DE-AC02-05CH11231 (LBNL) and DE-AC02-06CH11357 (ANL). This research used resources of ANL’s ATLAS facility, which is a DOE Office of Science User Facility. M.W.~acknowledges support by the National Research Foundation of South Africa under Grant No.~118846. We thank Alfredo Poves for enlightening discussions on the topic of this work and Birger Back for his help during the run. The authors also thank the operations staff of the ATLAS facility.  

\bibliography{references}

\end{document}